\documentclass[runningheads]{llncs}
\usepackage{graphicx}
\usepackage[hidelinks]{hyperref}
\usepackage{amssymb}

\begin{document}

\title{Multi-modal volumetric concept activation to explain detection and classification of metastatic prostate cancer on PSMA-PET/CT}

\titlerunning{Multi-modal volumetric concept activation}

\author{R.C.J. Kraaijveld\inst{1} \and M.E.P. Philippens\inst{2} \and W.S.C. Eppinga\inst{2} \and I.M. J{\"u}rgenliemk -Schulz\inst{2} \and K.G.A. Gilhuijs\inst{1} \and P.S. Kroon \inst{2} \and B.H.M. van der Velden\inst{1}}

\authorrunning{R.C.J. Kraaijveld et al.}
\institute{Image Sciences Institute, University Medical Center Utrecht, The Netherlands \and Department of Radiotherapy, University Medical Center Utrecht, The Netherlands}

\maketitle

\begin{abstract}
Explainable artificial intelligence (XAI) is increasingly used to analyze the behavior of neural networks. Concept activation uses human-interpretable concepts to explain neural network behavior. This study aimed at assessing the feasibility of regression concept activation to explain detection and classification of multi-modal volumetric data. 

Proof-of-concept was demonstrated in metastatic prostate cancer patients imaged with positron emission tomography/computed tomography (PET/CT). Multi-modal volumetric concept activation was used to provide global and local explanations.

Sensitivity was 80\% at 1.78 false positive per patient. Global explanations showed that detection focused on CT for anatomical location and on PET for its confidence in the detection. Local explanations showed promise to aid in distinguishing true positives from false positives. Hence, this study demonstrated feasibility to explain detection and classification of multi-modal volumetric data using regression concept activation.

\keywords{Explainable artificial intelligence \and XAI \and Interpretable deep learning \and Medical image analysis \and Prostate cancer \and PET/CT}
\end{abstract}

\section{Introduction}
\let\thefootnote\relax\footnote{This paper has been accepted as: Kraaijveld, R.C.J., Philippens, M.E.P., Eppinga, W.S.C., J{\"u}rgenliemk-Schulz, I.M., Gilhuijs, K.G.A., Kroon, P.S., van der Velden, B.H.M. ``Multi-modal volumetric concept activation to explain detection and classification of metastatic prostate cancer on PSMA-PET/CT." \textit{Medical Image Computing and Computer Assisted Intervention (MICCAI) workshop on Interpretability of Machine Intelligence in Medical Image Computing (iMIMIC)}, 2022.}

Deep learning has revolutionized medical image analysis. The neural networks used in deep learning typically consist of many layers connected via many nonlinear intertwined connections. Even if one was to inspect all these layers and connections, it is impossible to fully understand how the neural network reached its decision \cite{murdoch2019definitions}. Hence, deep learning is often regarded as a ``black box" \cite{murdoch2019definitions}. In high-stakes decision-making such as medical applications, this can have far-reaching consequences \cite{rudin2019stop}.

Medical experts have voiced their concern about this black box nature, and called for approaches to better understand the black box \cite{jia2020clinical}. Such approaches are commonly referred to as interpretable deep learning or explainable artificial intelligence (XAI) \cite{adadi2018peeking}. Visual explanation is the most frequently used XAI \cite{van2022explainable}. There is increasing evidence that the saliency maps that provide this visual explanation are to be used with caution \cite{adebayo2018sanity,arun2021assessing,eitel2019testing}. For example, they can be incorrect and not correspond to what the end-user expected from the explanation (i.e., low validity) or lack robustness \cite{van2022explainable}. Hence, such methods may not be as interpretable as desired. 

In response to ``uninterpretable" XAI, Kim et al. proposed to use human-interpretable concepts for explaining models (e.g. a neural network) \cite{kim2018interpretability}. Examples of such concepts are a spiculated tumor margin -- a sign of malignant breast cancer \cite{gilhuijs1998computerized} -- or the short axis of a metastatic lymph node in a prostate cancer patient, which has been related to patient prognosis \cite{meijer2013retrospective}. Using concepts, Kim et al. were able to test how much a concept influenced the decision of the model (i.e., concept activation) \cite{kim2018interpretability}.

Concept activation has been used in medical image analysis to explain classification techniques using binary concepts \cite{kim2018interpretability} -- such as the presence of micro-aneurysms in diabetic retinopathy -- and continuous concepts (i.e., regression concept activation) \cite{graziani2020concept} -- such as the area of nuclei in breast histopathology. To the best of our knowledge, the promise of concept activation has not yet been shown in detection, 3-dimensional volumetric data, or multi-modal data.

The aim of this study was to assess the feasibility of regression concept activation to explain detection and classification of multi-modal volumetric data. We demonstrated proof-of-concept in patients who had metastatic prostate cancer.

\section{Data}
A total of 88 consecutively included male patients with oligometastatic (i.e., five or less metastatic lymph nodes) prostate cancer from the University Medical Center Utrecht were analysed. All patients gave written informed consent and the study was approved by the local medical ethics committee \cite{werensteijn2021progression}. Median age was 71 years with an interquartile interval of 67--74 years.

\begin{figure}[h!]
    \centering
    \includegraphics[width=1.0\columnwidth]{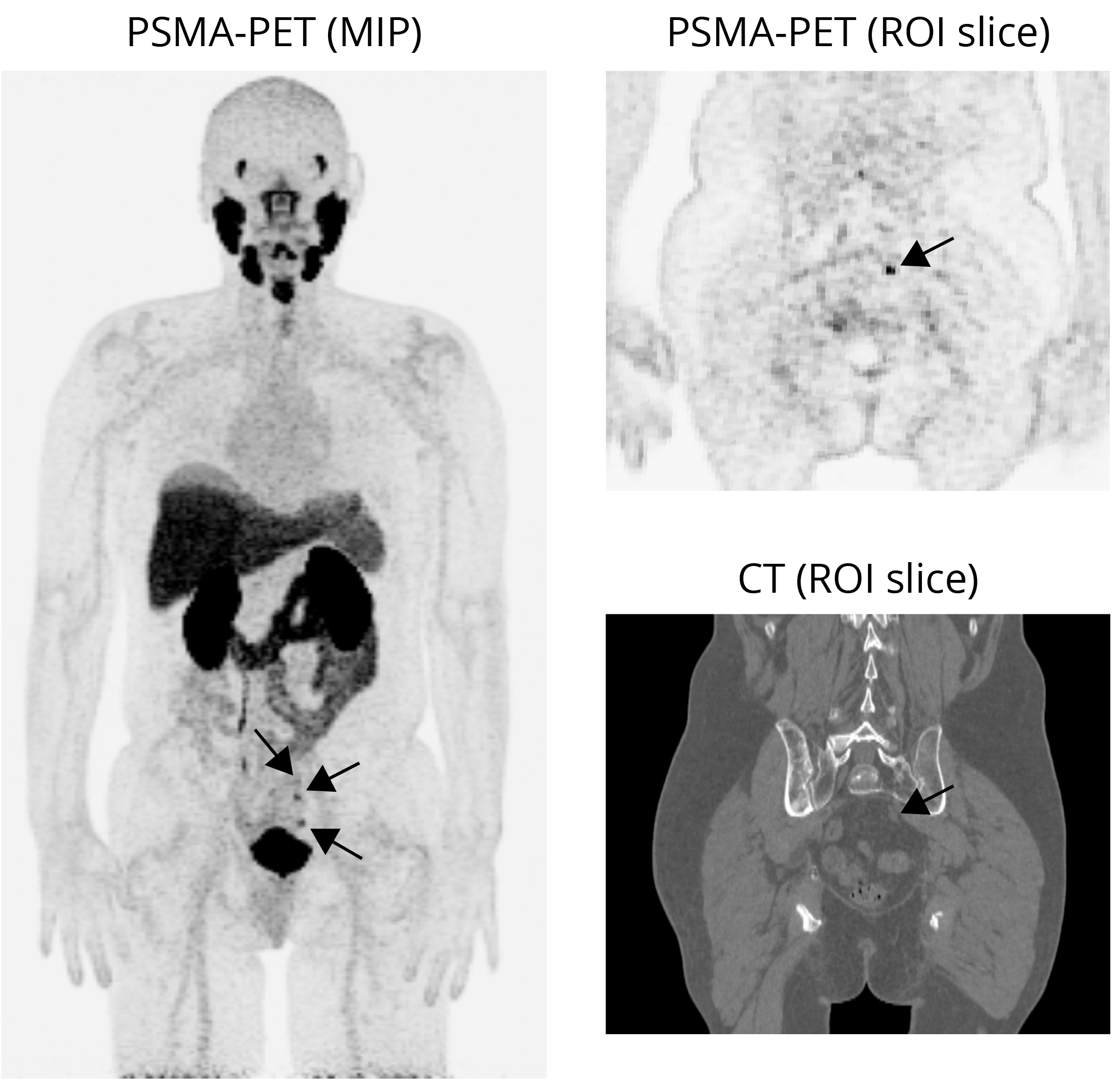}
    \caption{Example of a prostate cancer patient with three metastatic lymph nodes. Left: maximum intensity projection (MIP) of prostate-specific membrane antigen positron emission tomography (PSMA-PET) showing three metastatic lymph nodes. Right: region of interest (ROI) showing one of the metastatic lymph nodes on PSMA-PET and on computed tomography (CT).}
    \label{fig:fig1}
\end{figure}

Patients were imaged using \textsuperscript{68}Ga prostate-specific membrane antigen positron emission tomography and computed tomography (PSMA-PET/CT) (Figure \ref{fig:fig1}). The in-plane voxel size of the PET scans ranged from 1.5 mm\textsuperscript{2} to 4.1 mm\textsuperscript{2}, slice thickness ranged from 1.5 mm to 5.0 mm. The in-plane voxel size of the CT scans ranged from 0.84 mm\textsuperscript{2} to 1.4 mm\textsuperscript{2}, slice thickness was 2.0 mm.

Metastatic lymph nodes were delineated by a radiation oncologist in consensus with a nuclear medicine physician. Furthermore, lymph nodes were confirmed on magnetic resonance imaging.

\section{Method}
\begin{figure}[h!]
    \centering
    \includegraphics[width=1\columnwidth]{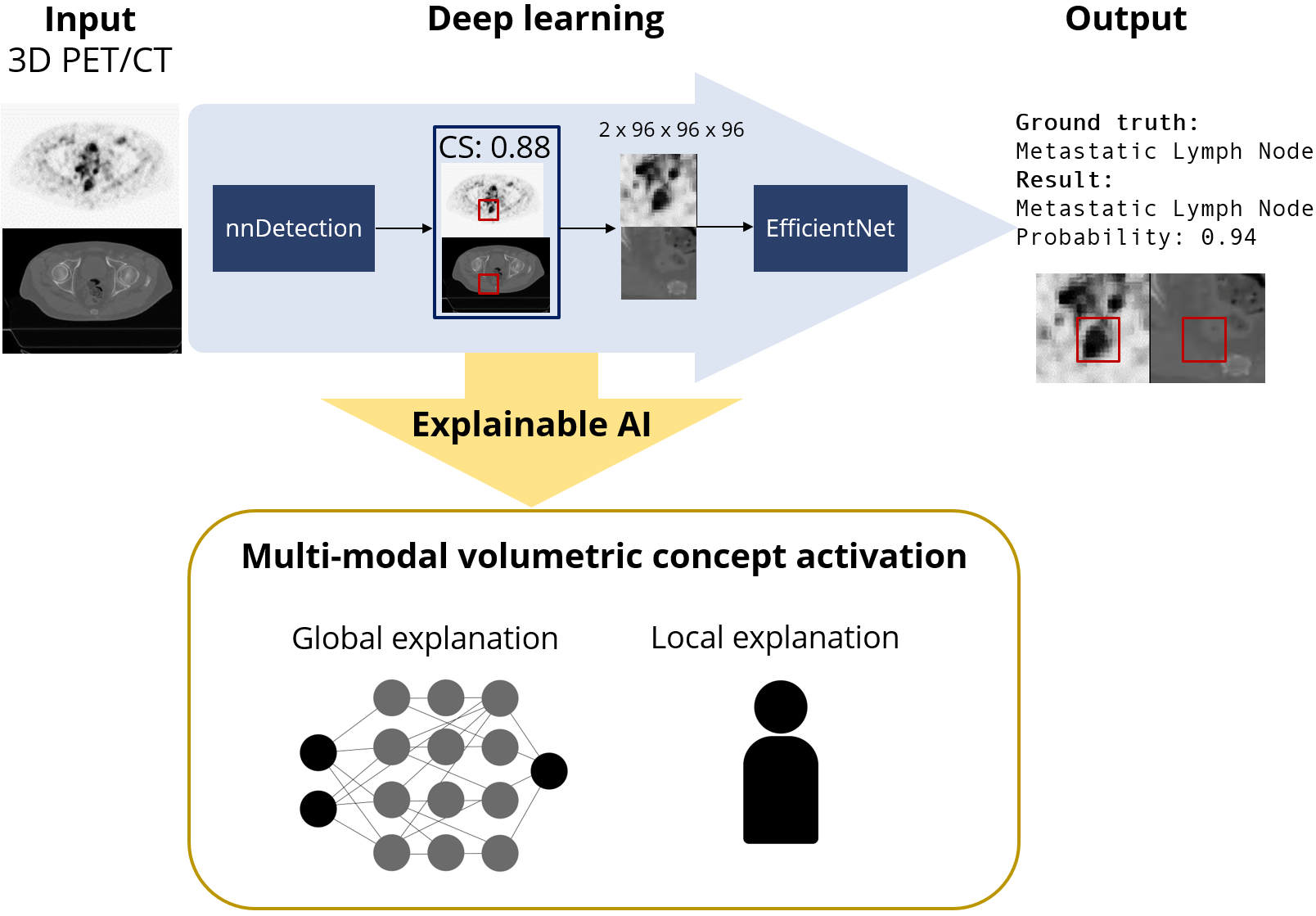}
    \caption{Schematic overview of the method. First, nnDetection detects metastatic lymph nodes on multi-modal volumetric positron emission tomography and computed tomography (PET/CT) images. These detections are then refined using EfficientNet. An XAI -- multi-modal volumetric concept activation -- is used to provide global and local explanations. CS = confidence score.}
    \label{fig:fig2}
\end{figure}

In short, we first detected the metastases and subsequently filtered out false positive detections at high sensitivity using classification. XAI was used on both the detection and the classification to provide global and local explanation (Figure \ref{fig:fig2}).

\subsection{Preprocessing}
PET scans were registered to the CT scans. Data was split into 70 patients for training/validation and 18 patients for testing. This resulted in 109 metastatic lymph nodes for training and 30 for testing.

\subsection{Detection}
nnDetection \cite{baumgartner2021nndetection} was used to detect the metastatic lymph nodes. Input to nnDetection were PET/CT images, output were 3D bounding boxes with corresponding intersection-over-union and confidence scores. Hyperparameters were optimized by nnDetection. 

The results of nnDetection were evaluated using Free-response Receiver Operating Characteristics. To ensure high metastatic lymph node detection rate, the intersection-over-union and confidence scores were thresholded at high sensitivity.

\subsection{Classification}
EfficientNet \cite{tan2019efficientnet} was used to subsequently filter out false positive detections by classifying bounding boxes originating from nnDetection. PET/CT volumes of $96\times96\times96$ (i.e., patches) were extracted. These patches were input to EfficientNet, output were binary classes representing whether there was a metastatic lymph node present or not. EfficientNet was trained using Adam optimizer and cross entropy loss. The initial learning rate was set as 0.001 and decreased step-wise by 0.10 every 5 epochs. EfficientNet was trained for 25 epochs with early-stopping. Augmentation included horizontal and vertical flipping, translation, scaling and rotation. Weighted random sampling was used to minimize the effect of class imbalance. 

The results of EfficientNet were evaluated using Receiver Operating Characteristics. To preserve true positives while reducing false positive that originated from nnDetection, the posterior probability per patch was thresholded at high sensitivity. 

\subsection{Explainable AI}
We provided explanations of both nnDetection and EfficientNet using volumetric regression concept attribution. 

Volumetric regression concept attribution yields global explanations, i.e., which concepts explain the overall behavior of the neural network, and local explanations, i.e., which concepts explain how the neural network came to a decision for a specific lymph node.

The concepts used in this study were extracted using PyRadiomics \cite{van2017computational}. This yields human-interpretable concepts per lymph node such as volume, circularity in 3D, and intensity on PET and CT, but also less interpretable concepts such as higher order texture features. The concepts were calculated from PET and CT, after applying masks which were automatically generated using an adaptive PET threshold of 40\% \cite{erdi1997segmentation,im2018current}.

Global explanations were provided using four measures that quantify volumetric regression concept attribution:

\begin{enumerate}
    \item Pearson's correlation coefficient $\rho$ was calculated between each feature and either the confidence scores in case of nnDetection or the posterior probability in case of EfficientNet. 
    
    \item The regression coefficient and regression concept vector were assessed per feature by fitting a linear model between layer activations and feature values. For each layer in the neural network, a regression coefficient can be quantified per concept, revealing the learning behavior of the neural network.   
    
    \item Sensitivity scores were calculated which indicate the influence of the concept on the outcome of the neural network result. 

    \item The bidirectional relevance was calculated for each concept by taking the product of the regression coefficient and the inverse of the coefficient of variation of the sensitivity scores.
\end{enumerate}

Local explanations were provided by comparing the sensitivity score of a concept per input image to the mean sensitivity of that concept. The difference between these sensitivity scores can be used as a similarity measure of that input image to an output class (e.g., metastatic lymph node).

\subsubsection{Computation:}
Deep learning was done in PyTorch 1.8 on an NVIDIA GeForce 2080Ti. Code will be available at  \url{https://github.com/basvandervelden/mmvca}.

\section{Results}
\subsection{Detection}

\begin{figure}[h!]
    \centering
    \includegraphics[width=0.7\columnwidth]{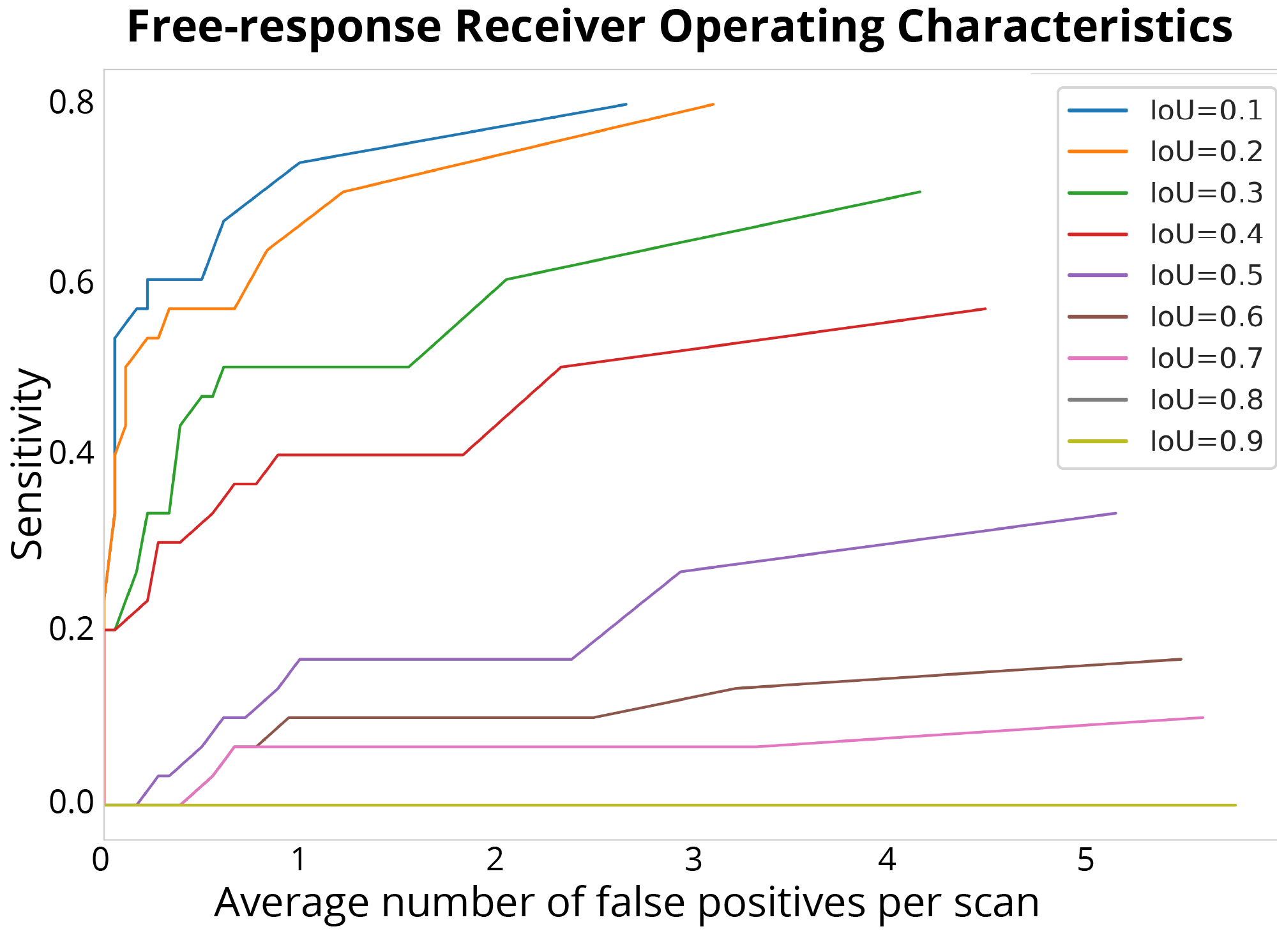}
    \caption{At an intersection-over-union (IoU) of 0.1, 0.80 sensitivity was obtained at 2.66 false positives per patient (top line). }
    \label{froc}
\end{figure}

At an intersection-over-union of 0.1, a sensitivity of 0.80 was obtained at an average of 2.66 false positive per patient (Figure \ref{froc}). In total, 24 out of 30 lymph nodes were detected at the cost of 48 false positives.

\subsection{Classification}
EfficientNet showed an additional reduction of 16 of the 48 false positives that originated from nnDetection (33\% reduction), while maintaining all true positives. Hence, the final amount of false positives per patient was 1.78.

\subsection{Explainable AI}

\subsubsection{Global explanations:}
\begin{figure}[h!]
    \centering
    \includegraphics[width=\linewidth]{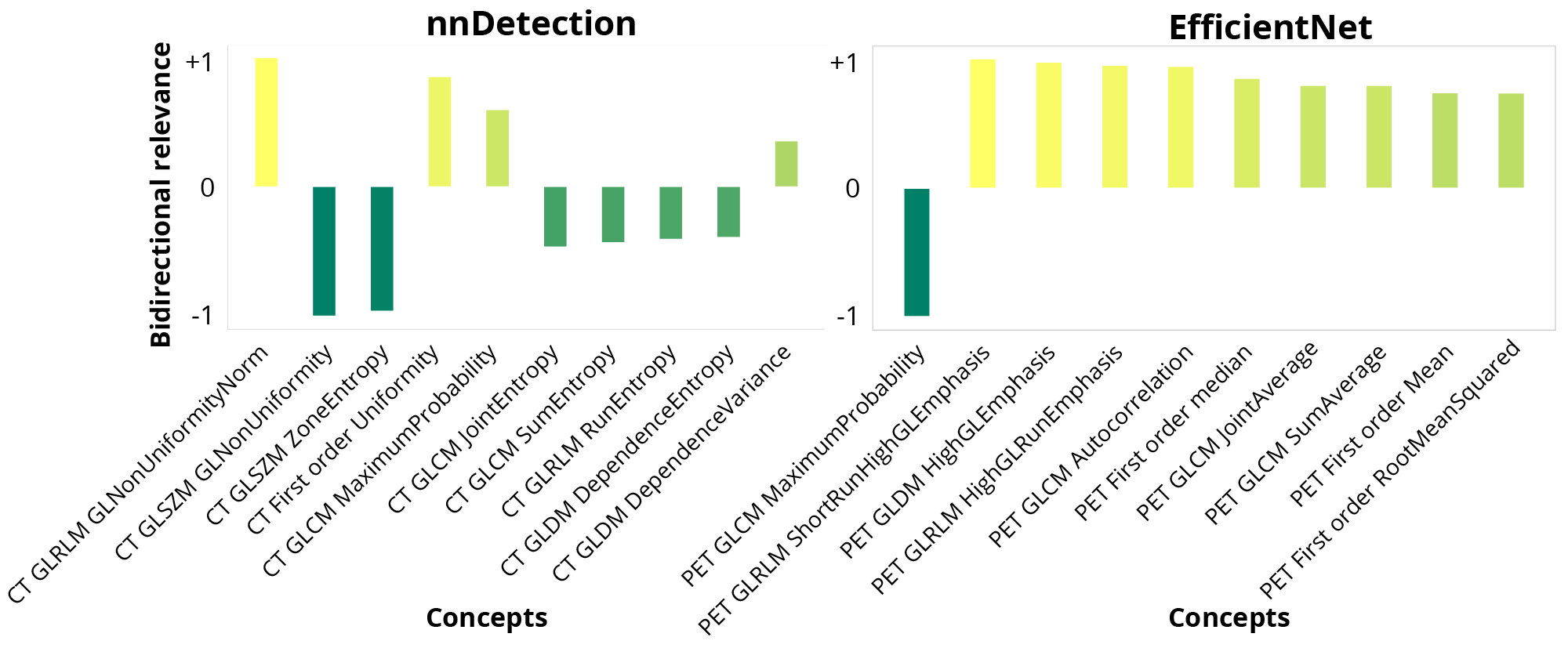}
    \caption{The top ten concepts with the highest bidirectional relevance originate from the computed tomography (CT) scan for nnDetection (left) and from the positron emission tomography (PET) scan for EfficientNet (right). GL = Gray level, Norm = normalized, GLRLM = Gray Level Run Length Matrix, GLSZM = Gray Level Size Zone Matrix, First order = First order statistics, GLCM = Gray Level Cooccurence Matrix, GLDM = Gray Level Dependence Matrix.}
    \label{BidirectionalRelevance}
\end{figure}

Table \ref{nndet_cor} shows the top ten concepts with the highest Pearson's correlation coefficient $\rho$ between the concepts and confidence scores of the bounding boxes from nnDetection. All these top ten concepts originate from the PET scan. Figure \ref{BidirectionalRelevance} shows the top ten bidirectional relevance scores for nnDetection. All these top ten concepts originate from the CT scan.

\begin{table}[h!]
\centering
\begin{tabular}{p{0.5\textwidth}p{0.12\textwidth}p{0.12\textwidth}}

\hline
\textbf{Concept}                                 & \textbf{$\rho$} & \textbf{P-value} \\ \hline \hline
PET GLCM DifferenceAverage           & 0.186     & $\leqq 0.001$                \\ \hline
PET GLCM DifferenceEntropy           & 0.185     & $\leqq 0.001$                \\ \hline
PET Firstorder Range                 & 0.185     & $\leqq 0.001$                 \\ \hline
PET GLSZM SizeZoneNonUniformity      & 0.176     & $\leqq 0.001$                 \\ \hline
PET Firstorder Maximum               & 0.175    & $\leqq 0.001$                 \\ \hline
PET GLRLM RunEntropy                 & 0.168     & $\leqq 0.001$                 \\ \hline
PET Firstorder Entropy               & 0.152     & $\leqq 0.001$                \\ \hline
PET GLCM SumEntropy                  & 0.148     & $\leqq 0.001$                \\ \hline
PET Firstorder MeanAbsoluteDeviation & 0.147     & $\leqq 0.001$                 \\ \hline
PET GLDM SmallDependenceEmphasis     & 0.140     & $\leqq 0.001$                 \\ \hline
\end{tabular}
\caption{All of the top ten correlations between concepts and the confidence scores of the bounding boxes originate from the positron emission tomography (PET) scan. GLCM = Gray Level Cooccurence Matrix, First order = First order statistics, GLSZM = Gray Level Size Zone Matrix, GLRLM = Gray Level Run Length Matrix, GLDM = Gray Level Dependence Matrix.}
\label{nndet_cor}
\end{table}

Table \ref{effnet_petct_cor} shows the top ten concepts with the highest Pearson's correlation coefficient $\rho$ between the concept and the posterior probability of a metastatic lymph node in the patch. Figure \ref{BidirectionalRelevance} shows which concepts influence the classification results the most. These top ten concepts for both XAI measures originate from the PET scan. 

\begin{table}[h!]
\centering

\begin{tabular}{p{0.5\textwidth}p{0.12\textwidth}p{0.12\textwidth}}
\hline
\textbf{Concept}                           & \textbf{$\rho$} & \textbf{p-value} \\ \hline \hline
PET First order Range           & 0.449      & $\leqq 0.001$                \\ \hline
PET GLCM SumAverage            & 0.444      & $\leqq 0.001$                \\ \hline
PET GLCM JointAverage          & 0.444      & $\leqq 0.001$               \\ \hline
PET First order Median          & 0.442      & $\leqq 0.001$                \\ \hline
PET First order Maximum         & 0.436      & $\leqq 0.001$                \\ \hline
PET First order Mean            & 0.430       & $\leqq 0.001$                \\ \hline
PET First order RootMeanSquared & 0.429      & $\leqq 0.001$                \\ \hline
PET GLCM MCC                   & 0.428      & $\leqq 0.001$               \\ \hline
PET First order 10Percentile    & 0.425      & $\leqq 0.001$                \\ \hline
PET First order 90Percentile    & 0.423      & $\leqq 0.001$                \\ \hline
\end{tabular}
\caption{All of the top ten correlations between concepts and the posterior probability of a metastatic lymph node in the patch originate from the positron emission tomography (PET) scan. First order = First order statistics, GLCM = Gray Level Cooccurence Matrix.}
\label{effnet_petct_cor}
\end{table}

\subsubsection{Local explanations:}

\begin{figure}[h!]
    \centering
    \includegraphics[width=\linewidth]{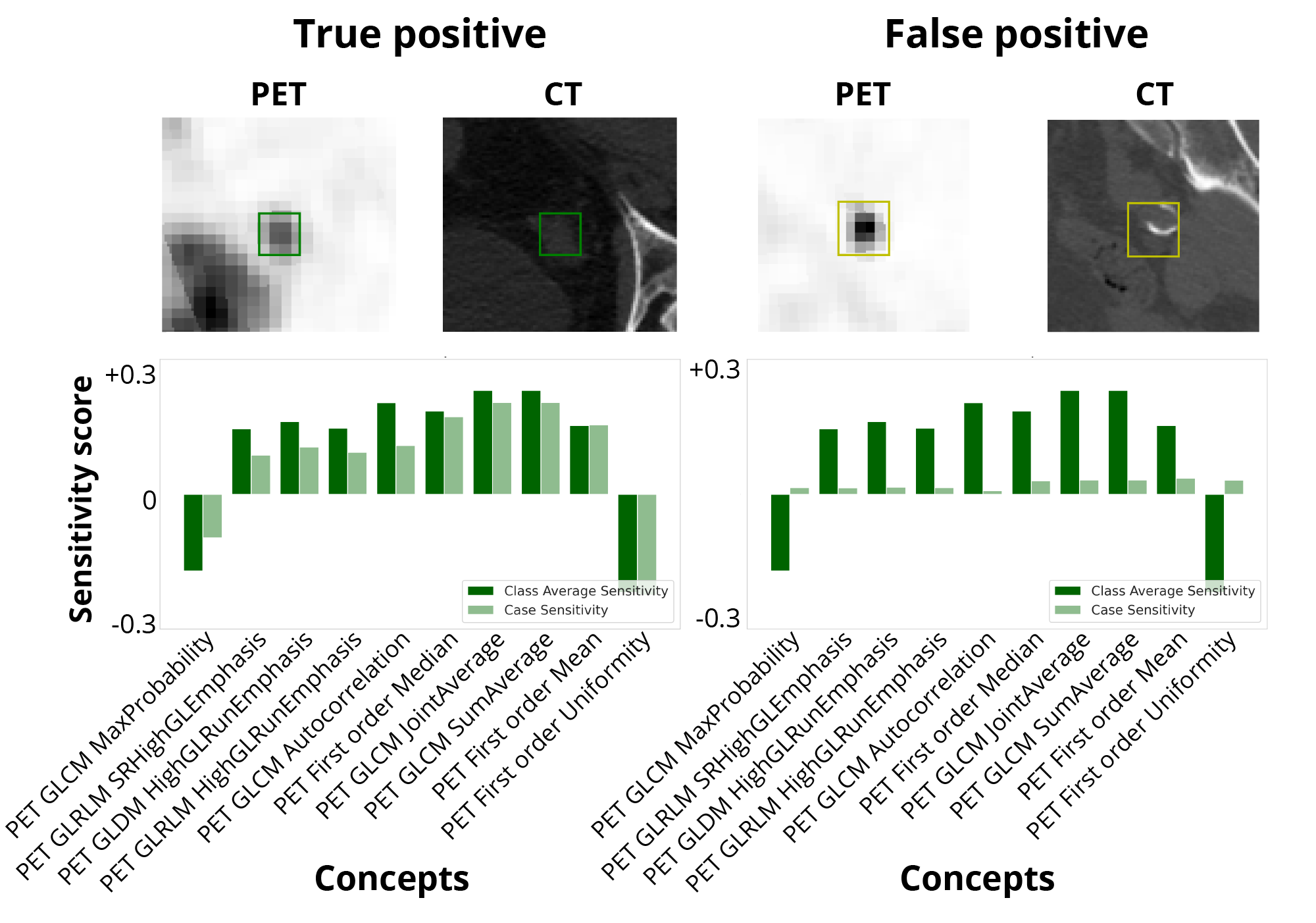}
    \caption{True positive (left) and false positive finding (right) with their corresponding local explanation underneath. It can be seen that the sensitivity scores of the left PET/CT patch reflects the class sensitivity scores. In the right PET/CT patch the sensitivity scores differ substantially from the class sensitivity scores. Hence, this local explanation can give an extra confirmation to the physician to rule this a false positive. 
    GLCM = Gray Level Cooccurence Matrix, GLRLM = Gray Level Run Length Matrix, GLDM = Gray Level Dependence Matrix, First order = First order statistics.}
    \label{LocalExplanation}
\end{figure}

Figure \ref{LocalExplanation} shows how the local explanations can be used by a physician. Each case was ranked according to its similarity with a metastatic lymph node and its top ten concepts. 

To further investigate the six undetected lymph nodes from nnDetection, we also evaluated these in a post hoc analysis with EfficientNet. Four of the six (66\%) false negatives were correctly classified as a lymph node. Local explanations showed that the two incorrectly classified lymph nodes had low similarity with the class metastatic lymph node, according to the top ten concepts.

\section{Discussion}
This study showed feasibility of regression concept activation to explain detection and classification of multi-modal volumetric data. In 88 oligometastatic prostate cancer patients, our method was able to provide realistic global and local explanations.

The global explanations for nnDetection yielded plausible results. Confidence scores of nnDetection's bounding boxes were all positively correlated with concepts from the PET scan, whereas the concepts that influenced the position of the bounding boxes came from the CT scan. In other words, the CT scan provides detailed anatomical information explaining in which region of the patient lymph nodes could be present, whereas the PET scan influences how confident the network is that the detection is actually a metastatic lymph node. Since PSMA-PET is designed for this specific goal, these explanations are plausible.

The global explanations for EfficientNet also yielded plausible results. The posterior probability whether a metastatic lymph node was present in a patch was mostly correlated with concepts from the PET scan. This again makes sense, since the volume of interest was already narrowed down, making the anatomical information from the CT scan less important in this part of the analysis.

Local explanations were aimed at providing a framework for physicians to evaluate on an individual lesion basis how the algorithm came to its conclusion, and whether they trust the algorithms decision. This has potential for decision support in the more difficult lesion in which the physician is potentially unsure.

This study has some limitations. Firstly, nnDetection misses six metastatic lymph nodes, leading to a sensitivity of 80\%. This is, however, similar to sensitivities reported in literature \cite{kim2019diagnostic}. The local explanations yielded insight into why these six false negative lymph nodes were not detected: Their concepts showed a large contrast with for example the detected lymph nodes. By taking this into account, in future work, the explanations can be used to further optimize the neural network \cite{lund2022leveraging,mahapatra2022self}. Secondly, we did not evaluate our explanations with end-users such as radiation oncologists. Future work should evaluate these explanations with intended end-users, i.e., application-grounded evaluation \cite{doshi2017towards}. Lastly, we demonstrate our approach in a single center study population. Larger validation would be desired in future research.

\section{Conclusion}
To conclude, we showed that it is feasible to explain detection and classification of multi-modal volumetric data using regression concept activation.


\bibliographystyle{splncs04}
\bibliography{bib}

\end{document}